\documentclass[twocolumn,superscriptaddress,secnumarabic,
amssymb,amsmath,nobibnotes,aps,prd,showkeys,showpacs,nofootinbib]{revtex4}%
\usepackage{graphicx}
\usepackage{epsf}
\usepackage{bm}
\usepackage{amsmath}
\usepackage{amsfonts}
\usepackage{amssymb}
\usepackage{epstopdf}
\usepackage{natbib}%
\usepackage{rotating}
\usepackage{pdflscape}
\usepackage{adjustbox}
\providecommand{\U}[1]{\protect\rule{.1in}{.1in}}
\newcommand{\be}{\begin{equation}}
\newcommand{\ee}{\end{equation}}

\newcommand{\mincir}{\raise
-3.truept\hbox{\rlap{\hbox{$\sim$}}\raise4.truept\hbox{$<$}\ }}
\newcommand{\magcir}{\raise
-3.truept\hbox{\rlap{\hbox{$\sim$}}\raise4.truept\hbox{$>$}\ }}

\usepackage{color}

\begin{document}
\title{Cosmological consequences of an adiabatic matter creation process}
\author{Rafael C. Nunes}
\email{nunes@ecm.ub.edu}
\affiliation{CAPES Foundation, Ministry of Education of Brazil, Bras\'ilia - DF 70040-020, Brazil}
\author{Supriya Pan}
\email{span@iiserkol.ac.in}
\affiliation{Department of Physical Sciences, Indian Institute of Science Education and Research -- Kolkata, Mohanpur -- 741246, West Bengal, India}
\keywords{cosmological parameters - dark energy - equation of state - dark matter - cosmic background radiation}
\pacs{98.80.-k, 95.35.+d, 95.36.+x, 98.80.Es.}
\begin{abstract}

In this paper we investigate the cosmological consequences of a continuous matter creation associated with the production of particles by the gravitational field acting on the quantum vacuum. To illustrate this, three phenomenological models are considered. An equivalent scalar field description is presented for each models. The effects on the cosmic microwave background power spectrum are analyzed for the first time in the context of adiabatic matter creation cosmology. Further, we introduce a model independent treatment, $Om$, which depends only on the Hubble expansion rate and the cosmological redshift to distinguish any cosmological model from $\Lambda$CDM by providing a null test for the cosmological constant, meaning that, for any two redshifts $z_1$, $z_2$, $Om (z)$ is same, i.e. $Om (z_1)- Om (z_2)= 0$. Also, this diagnostic can differentiate between several cosmological models by indicating their quintessential/ phantom behavior without knowing the accurate value of the matter density, and the present value of the Hubble parameter. For our models, we find that particle production rate is inversely proportional to $Om$. Finally, the validity of the generalized second law of thermodynamics bounded by the apparent horizon has been examined.

\end{abstract}

\maketitle
\section{Introduction}
\label{sec:intro}

The present accelerated expansion of the universe is one of the biggest and fascinating cosmic puzzles since its discovery
\cite{Riess98,Perlmutter99}. Nowadays, we have many independent observational results
\cite{Percival2001,Spergel2003,Tegmark04,Eisenstein05,Komatsu11} that confirmed this claim.
In the context of relativistic cosmology, an accelerating regime is obtained by assuming an existence of some dark energy (DE) component, an exotic fluid endowed with negative pressure that occupies about 3/4 of current energy density of the universe. The cosmological constant, interpreted as the density of energy associated with the quantum vacuum, has become the main candidate for DE. Such cosmological model is the so called $\Lambda$ - cold dark matter ($\Lambda$CDM) model, and presents two serious problems, namely, the cosmological constant problem, and the cosmic coincidence problem (for a review on these issues, see \cite{Weinberg, SS2000,Padmanabhan2003}. Because of these problems, several alternative models have been proposed to explain the late-time accelerated universe \cite{AT}.\newline

However, according to recent observations \cite{Planck2015,Planck2013,Rest,Xia,Cheng,Shafer} the equation of state (EoS) of dark energy, $w$, has crossed the ``$-1$'' boundary. In other words, our universe has a slight phantom nature, hence, we need something else which can go beyond `$-1$'.
A natural candidate to realize this  $w < -1$, scenario is to introduce a phantom scalar field whose energy density ($\rho_{\phi}$)
and pressure ($p_{\phi}$) take the forms, $\rho_{\phi}= -\dot{\phi}^2/2+ V (\phi)$, and $p_{\phi}= -\dot{\phi}^2/2- V (\phi)$, respectively. Hence, the EoS, $w_{\phi}= p_{\phi}/\rho_{\phi}< -1$. Since the kinetic part contains the wrong sign \cite{Caldwell},
phantom scalar fields lead to some instabilities at the classical
and quantum levels \cite{Carroll,Cline}. Moreover, it has been noticed that, this kind of fields suffer from other serious theoretical problems \cite{Hsu,Sbisa,Dabrowski}. It is therefore natural to look for
alternative ways to drive the accelerated expansion of the universe up to the phantom region without
any of the aforementioned difficulties. A viable choice which could be an alternative to
both dark energy and gravity modified models, is the matter creation models by
gravitational field, where this kind of models can mimic $\Lambda$CDM cosmology \cite{ccdm_model}. Note that, matter creation
in cosmology is not a new idea, rather it has a long history since its introduction 
by Parker and collaborators \cite{Parker}.\newline

Parker assumed that, the material content of the universe may have lead its origin from the continuous creation of radiation and matter from the gravitational field of the expanding universe acting on the quantum vacuum. This continuous matter creation does not depend on the theory of gravity. Further, the produced particles have some well defined properties --- they gain
mass, momentum and energy from the time-evolving gravitational background which converts curvature into particles by 
a pumping mechanism, as if the gravity pumps on the curvature to produce particles.\newline

%

Prigogine  et al. \cite{Prigogine} compelled matter creation into the Einstein's field equations,
in a consistent way by introducing in the usual balance equation for the number density of particles, $ (n\, u^{\alpha})_{; \alpha}=0$,
a source term in the right hand side that accounts for the production of particles as

\begin{eqnarray}
(n\, u^{\alpha})_{; \alpha} = n \Gamma .
\label{Eq:nbalance}
\end{eqnarray}
Here, $n$ is the number density of the fluid particles, $u^{\alpha}$ is the four-velocity vector of the created particles which is normalized, so that $ u^{\alpha}\, u_{\alpha} = 1$, and $\Gamma$ denotes the
particle production rate. Note that, in the radiation dominated era, the production of radiation particles vanished (Parker's theorem) \cite{Parker-Toms}. Taking into account the second law of thermodynamics in the discussion, by means of a covariant formalism, it has been found that, Eq. (\ref{Eq:nbalance}) can lead some extra pressure quantity, which is directly related to the particle creation rate $\Gamma$, and hence they termed as ``creation pressure'' \cite{LCW, LGA, mnras-winfried}, which has further been confirmed by using relativistic kinetic theory \cite{cqg-triginer,Lima-baranov}.
Since the entropy flux vector of matter, $n \sigma u^{\alpha}$, where $\sigma$ denotes the entropy per particle, must fulfill the second law of thermodynamics $(n \sigma u^{\alpha})_{;\alpha} \geq 0$, therefore we have the constraint $\Gamma \geq 0$ on the particle creation rate.
Hence, matter creation entered into the cosmological domain and confirmed its place \cite{ccdm_model}.\newline

Recently, Nunes and Pav\'{o}n \cite{NP2015} showed that, the matter creation models can realize the phantom universe without the need of invoking any phantom fields \cite{Caldwell}. Motivated by the models introduced in \cite{NP2015}, we establish a canonical scalar field description equivalent to matter creation and study its behavior with the expansion of the universe. Further, we analyze the effects of this matter creation on the cosmic microwave background spectrum compared to those predicted by the $\Lambda$CDM model. Then we introduce a model independent treatment for matter creation models, which is generally used to filter several cosmological models from $\Lambda$CDM by providing a null test, where the test also tells whether the cosmological model has a phantom or quintessence behavior. It is interesting to note that, we do not need to know the value of the present day Hubble parameter and the accurate value of the matter density for such test. In addition to that, any particle creation model, must agree with thermodynamics, more specifically, with the generalized second law of thermodynamics.\newline

This paper is organized as follows: The next section \ref{sec:models} gives a brief overview of phenomenological matter creation in cosmology, and introduces the models. Section \ref{sec:SF} presents an equivalent scalar field description of the models. Section \ref{sec:cmb} analyses the effects on the CMB power spectra coming from the matter creation models, and also compares those effects with respect to the standard cosmological model $\Lambda$CDM. Section \ref{sec:OM} presents a model independent treatment, called $Om$ diagnostic, for our models. In section \ref{sec:thermodynamics}, we discuss the generalized second law of thermodynamics (GSLT) bounded by apparent horizon. The concluding section summarizes and gives comments on our findings. As usual, a subindex zero attached to any quantity means that it must be evaluated at present time. We note that, throughout the text, we have used matter creation, particle creation, particle production synonymously.

\section{Cosmology of particle creation models}
\label{sec:models}

\noindent Let us consider a spatially flat Friedmann-Lema\^{\i}tre- Robertson-Walker (FLRW) universe

\begin{eqnarray}\label{FLRW}
ds^2&=& -dt^2+ a^2 (t) (dx^2+dy^2+dz^2),
\end{eqnarray}	
where $a (t)$ is the scale factor of the FLRW universe.
The Einstein's field equations for a perfect fluid endowed with an adiabatic particle production can be written as

\begin{eqnarray}
3 H^2&=& 8 \pi G \rho,\label{Friedmann1}\\
2 \dot{H}+ 3 H^2&=& - 8 \pi G (p+p_c),\label{Friedmann2}
\end{eqnarray}
where an overdot means differentiation with respect to the cosmic time; $H = \dot{a}/a$, is the Hubble rate; $\rho$, $p$ are respectively the energy density and pressure of the matter content. As is well-known, this $p_c$ is given by \cite{LCW, CLW, LG, LGA, mnras-winfried, Zimdahl}

\begin{eqnarray}\label{pressure_creation}
p_c= - \frac{\rho \, + \, p}{3H}\,
\Gamma \, .
\end{eqnarray}
Therefore, if $p_{c}$ is negative, it may drive the accelerated expansion of the universe.
It has been shown that, the production of ordinary particles is much limited by the tight constraints imposed
by local gravity measurements \cite{plb_ellis, peebles2003, hagiwara2002}, and radiation has practically no
impact on the late-time accelerated expansion of our universe. Therefore,
we assume that the produced particles are just the cold dark matter particles.

In this study, we consider that the energy density splits in three components, baryons, cold dark matter, and the quantum vacuum, i.e. $\rho= \rho_b+ \rho_{dm}+ \rho_{\Lambda}$. Now, as the cold dark matter particles are created from the gravitational field, hence Eq. (\ref{pressure_creation}) takes the form

\begin{align}\label{new-pressure-creation}
p_c & = - \frac{\rho_{dm}}{3H} \Gamma~~~~~\Longleftrightarrow ~~~~~w_c \equiv \frac{p_c}{\rho_{dm}}= -\frac{\Gamma}{3H}
\end{align}
where $w_c$ is defined as the equation of state due to the cold dark matter creation by this gravitational field, and this is negative for an expanding universe, i.e. for $H> 0$. The conservation equation for this particle creation becomes

\begin{eqnarray}
\label{dark_matter_evolution} \dot{\rho}_{dm} \,  + \, 3H \rho_{dm} = \rho_{dm} \, \Gamma.
\end{eqnarray}
In deriving (\ref{dark_matter_evolution}), we have used Eq. (\ref{Eq:nbalance})
specialized to dark matter particles, and the relation $\rho_{dm} = n_{dm} \, m$, where $m$ stands for the rest mass
of a typical dark matter particle, and $n_{dm}$ is the number of dark matter particles.
Since baryons are neither created nor destroyed, the baryonic sector satisfies the
following conservation law
\begin{eqnarray}\label{baryon-conservation}
\dot{\rho}_{b} \, + \, 3H \rho_{b} = 0.
\end{eqnarray}
On the other hand, the cosmological constant \cite{Weinberg, SS2000, Padmanabhan2003} interpreted as the density of energy associated to the quantum vacuum is constant, i.e. $\rho_{\Lambda} = {\rm constant}$. Thus, Friedmann equation (\ref{Friedmann1}) for this scenario can simply be expressed as

\begin{eqnarray}
\frac{H^2(a)}{H^2_0}=\Omega_{b0}\, a^{-3} \, +\, \Omega_{dm0}  a^{-3} \Big( \exp{\int_1^a \frac{da}{a}\frac{\Gamma}{H}} \Big) +
 \, \Omega_{\Lambda0},
 \label{Eq:Hz}
\end{eqnarray}
where $\Omega_{b0}$, $\Omega_{dm0}$, $\Omega_{\Lambda0}$ are respectively the density parameters for baryons, cold dark matter and vacuum energy, constrained by the relation $\Omega_{b0}+ \Omega_{dm0}+ \Omega_{\Lambda0}= 1$.

From Eq. (\ref{Eq:Hz}), the dynamics of this model can well be understood once $\Gamma$ is prescribed.
Unfortunately, the exact functional form for $\Gamma$ is very difficult to obtain  before the nature of created cold
dark matter particles is known. Hence, we must resort to phenomenological models
for $\Gamma$, and constrain the dynamics with the observational data. In order to achieve our goal,
we use three parameterizations of $\Gamma$ found in Ref. \cite{NP2015}

\begin{eqnarray}\label{proposta1}
\qquad \qquad \Gamma = 3 \beta H  \qquad \qquad
\qquad \; \; \; {\rm (Model \,\,\,\,\ I)},
\end{eqnarray}
\begin{eqnarray}\label{proposta2}
\Gamma = 3 \beta \, H \,  [5-5 \tanh (10-12a)]
\qquad {\rm (Model \,\,\,\,\ II)},
\end{eqnarray}
and
\begin{eqnarray}
\label{proposta3} \Gamma = 3 \beta \, H \, [5-5 \tanh (12a-10)]
\qquad {\rm (Model \,\,\,\,\ III) },
\end{eqnarray}
where $\beta> 0$ is a free parameter of the model characterizing the particle production process, $\beta=0$ means that, there is no matter creation, otherwise there is production of particles. If $\beta$ is so high then the matter production rate must be so high. In all cases, $\Gamma/3H \leq 1$.
Now, solving the conservation equation (\ref{dark_matter_evolution}) for the three particle creation models, we find that

\begin{align}\label{new-conservation}
\rho_{dm} & = \rho_{dm0} a^{-3} \exp\left(3 \beta \int_{1}^{a} \frac{g (\tilde{a})}{\tilde{a}} da\right)
\end{align}
where $\rho_{dm0}$, is the present energy density of the cold dark matter, and $g (a)= \Gamma/3H$,
which for all three models can explicitly be given by

\begin{eqnarray}
g(a) &=& 1,~~~(\mbox{Model I})\label{gaIa}\\
g(a)&=& 5-5 \tanh(10-12 a),~~(\mbox{Model II})\label{gaIIa}\\
g(a)&=& 5-5 \tanh(12 a-10).~~(\mbox{Model III})\label{gaIIIa}
\end{eqnarray}

For model I, $\rho_{dm} \propto a^{-3 (1-\beta)} \sim a^{-3}$, for very small value of the parameter $\beta$, while for models II, III, $g(a)/a \longrightarrow 0$, as $a \longrightarrow  \infty$, hence the creation of dark matter particles does not exceed the standard evolution $\rho_{dm} \propto a^{-3}$ in future evolution of the universe.
It is worthy to mention that, the scenario proposed in \cite{NP2015} takes us to the phantom region, i.e. the effective equation of state (EoS): $w_{eff}= w_{\Lambda}+ w_c < -1$, where $w_c$ is the EoS associated to the creation pressure (see Eq. (\ref{new-pressure-creation})), and $w_{\Lambda} (= - 1)$ is the EoS of the quantum vacuum. So, naturally, the need of phantom
fields will be weaken \cite{Caldwell, Carroll, Cline, Hsu, Sbisa, Dabrowski}. From a joint analysis of data Supernova type Ia, gamma ray bursts, baryon acoustic oscillations, and the Hubble rate, it was obtained that, $w_{eff}(z=0) = -1.073^{+0.034}_{-0.035}$, $-1.155^{+0.076}_{-0.080}$, and $-1.002^{+0.001}_{-0.001}$
for models I, II, and III, respectively at 1$\sigma$ confidence level (see \cite{NP2015}). Further, we note that, the current fractional densities of matter (baryons $+$ dark matter) is constrained as $\Omega_{m0}=\Omega_{b0} + \Omega_{dm0} \simeq 0.30$, for all three models.

\section{Canonical scalar field description}
\label{sec:SF}

Here, we obtain an equivalent field theoretic description for the particle creation models.
To do this, we replace the energy density ($\rho_{dm}$) and the creation pressure ($p_c$) for the particle creation models
to the corresponding energy density ($\rho_{\varphi}$) and the pressure ($p_{\varphi}$) for a canonical scalar field
$\varphi$ with potential $V(\varphi)$. On the background of the flat FLRW universe given in Eq. (\ref{FLRW}), and restricting
the field as $\varphi(t, x) = \varphi(t)$, i.e. to be homogeneous, we can write

\begin{eqnarray}\label{scalar-field}
\rho_{dm} = \rho_{\varphi}= \frac{1}{2} \dot{\varphi}^2+ V (\phi), \,\,\,\,\,\,\,\,\,\,
p_{c}= p_{\varphi}= \frac{1}{2} \dot{\varphi}^2- V (\varphi).
\end{eqnarray}
By combining Eqs. (\ref{scalar-field}), we obtain
\begin{eqnarray}
\label{phi}
\dot{\varphi}^2 = (1+ w_c)\rho_{dm},
\end{eqnarray}
and
\begin{eqnarray}
\label{V}
 V(\varphi) = \frac{1}{2} (1 - w_c)\rho_{dm},
\end{eqnarray}
or, in terms of $z$, we have

\begin{eqnarray}
\label{phi2}
 \dot{\varphi} = \frac{d \varphi}{dz} \dot{z} = - \frac{d \varphi}{ dz} (1+z) H(z),
\end{eqnarray}
so that

\begin{eqnarray}
\label{phi3}
 \frac{d \varphi}{ dz} = \mp \frac{1}{(1+z)H(z)} \sqrt{(1+w_c) \rho_{dm}},
\end{eqnarray}
where the negative (positive) sign stands for $\dot{\varphi} > 0$ ( $\dot{\varphi} < 0$). In order to allow the
field rolls down the potential as the universe expands, we choose the negative sign. In the above equations, $w_c$,
is the EoS associated to particle creation obtained from 
(\ref{proposta1}) -- (\ref{proposta3}),
for the models I, II, and III as follows:

\begin{eqnarray}
w_c &=& -\beta,\label{w_c_1}\\
w_c &=& -\beta [5-5 \tanh (10-12a)]\label{w_c_2}\\
w_c &=& -\beta [5-5 \tanh (12a-10)]\label{w_c_3}.
\end{eqnarray}


By defining $\phi = \sqrt{\frac{8\pi G}{3}}\,\,\varphi$, and $\Omega_V = V(\varphi)/\rho_{c,0}$, we have the following results for three models:
\\


%

\begin{figure}
 	\includegraphics[width=8.5cm, height=3.5cm]{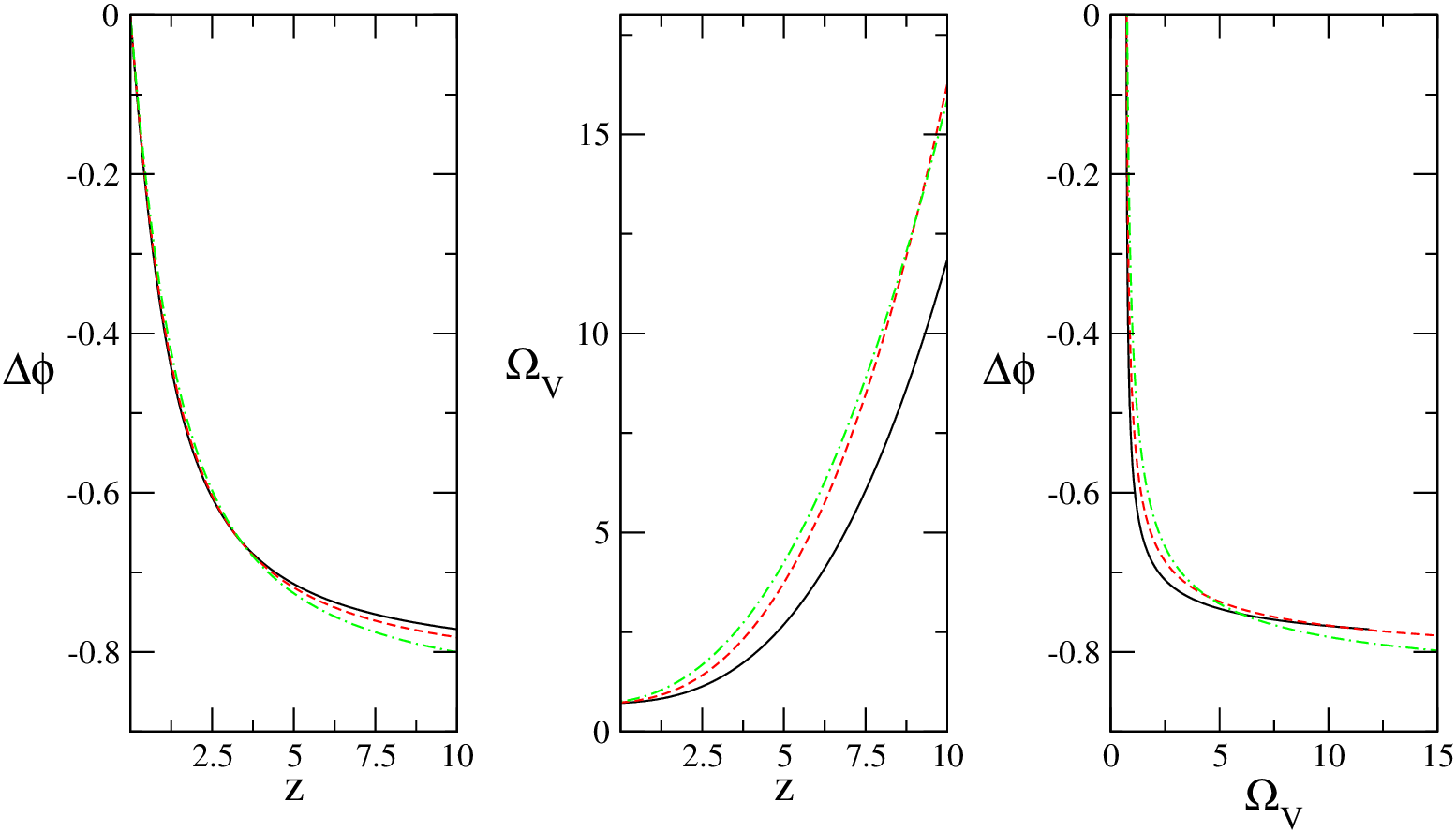}
     \caption{An equivalent scalar field description for model I (see Eq. \ref{proposta1}) is presented. The left panel shows
the variation of the scalar field, $\Delta \phi$, with respect to the redshift $z$. Middle panel: variation of the potential, $\Omega_V$, over the redshift. Right panel: dependence of the potential on the variation of $\phi$. Solid (black), dashed (red), and dot-dashed (green), stand for $\beta = 0.05, 0.1, 0.2$, respectively. While drawing the graphs, we  have taken $\Omega_{b0} = 0.05$, $\Omega_{dm0}=0.24$.}\vskip 0.5 cm
     \label{campo_model1}
\end{figure}

\begin{figure}
\includegraphics[width=8.5cm, height=3.5cm]{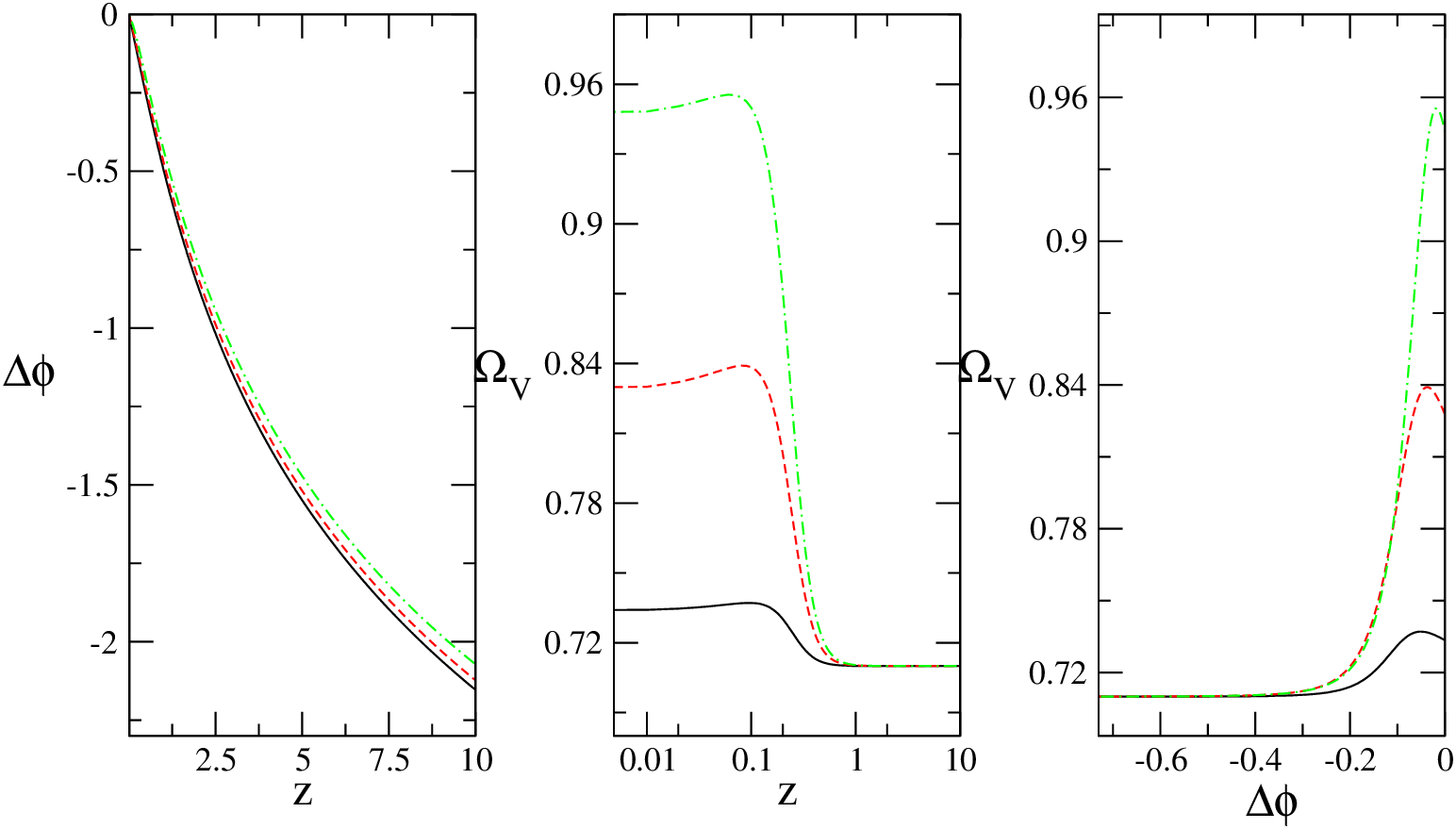}
\caption{This stands for an equivalent scalar field prescription for model II (see Eq. \ref{proposta2}).
Left panel: $\Delta \phi$ versus $z$. Middle panel: $\Omega_V$ versus $z$. Right panel: $\Omega_V$ versus $\Delta \phi$. Solid (black), dashed (red), and dot-dashed (green), stand for $\beta = 0.05, 0.1, 0.2$, respectively. For drawing the graphs, we  have taken $\Omega_{b0} = 0.05$, $\Omega_{dm0}=0.24$.}
\label{campo_model2}
\end{figure}

Model I:

\begin{eqnarray}
\label{Delta_model1}
  \Delta \phi = \phi - \phi_{0} = - \int_0^z \frac{dz'}{(1+z')\eta_1(z')} \Bigl[ (1+w_c) \times \nonumber \\
   \Omega_{dm0}(1+z')^{3(1-\beta)} \Bigr]^{\frac{1}{2}},
\end{eqnarray}
and
\begin{eqnarray}
 \Omega_V (z) = \frac{1}{2}(1-w_c) \Omega_{dm0}(1+z)^{3(1 -\beta)},
\end{eqnarray}
where $\eta_1(z)=[\Omega_{b0}(1+z)^3 + \Omega_{dm0}(1+z)^{3(1-\beta)} + \Omega_{\Lambda0}]^{1/2}$.
\\

Models II and III:\\

In this case, we find

\begin{eqnarray}
 \Delta \phi = \phi - \phi_{0} = - \int_0^z \frac{dz'}{(1+z')\eta_2(z')} \Bigl[ (1+ w_c) \times \nonumber\\
 \Omega_{dm0}(1+z')^3 \exp\left(-3 \beta \int_0^z  \frac{\psi (z')}{(1+z')}d z' \right)\Bigr]^{\frac{1}{2}},
\end{eqnarray}
and

\begin{eqnarray}
\Omega_V (z)= \frac{1}{2}(1-w_c) \Bigl[
\Omega_{dm0} (1+z)^3 \times \nonumber\\ \exp\left(-3 \beta \int_0^z  \frac{\psi (z')}{(1+z')} dz' \right) + \Omega_{\Lambda0}\Bigr],
\end{eqnarray}
where $\psi(z)$ is given by

\begin{eqnarray}
\label{g_a}
\psi(z)&=& 5-5 \tanh \Big(10-\frac{12}{1+z} \Big),~~~(\mbox{Model II})\label{gaII}\\
\psi(z)&=& 5-5 \tanh \Big(\frac{12}{1+z} -10 \Big),~~~(\mbox{Model III})\label{gaIII}
\end{eqnarray}
and

\begin{eqnarray}
\label{eta_23}
\eta_2(z)= \Bigl[ \Omega_{b0} (1+z)^3 + \nonumber\\
\Omega_{dm0} (1+z)^3 \exp\left(-3 \beta \int_0^z  \frac{\psi (z')}{(1+ z)}d z \right) + \Omega_{\Lambda0} \Bigr]^{\frac{1}{2}}.
\end{eqnarray}

We apply numerical simulation in Eqs. (\ref{Delta_model1}) -- (\ref{eta_23}) to obtain the complete scalar field description
for the parametric models proposed in this work.\\

Figure \ref{campo_model1} depicts an equivalent field theoretic description for model I which shows the
evolution of the scalar field with respect to the cosmological redshift (see left panel); potential versus redshift (middle panel);
and the scalar potential with respect to the field itself, for some values of $\beta$.
Note that, for $z \lesssim 2$, different values for the parameter $\beta$ has little influence on
the variation of field, $\Delta \phi$, and on the scalar potential, $V (\varphi)$. Thus, it is expected
that different values for the particle production rate, do not create any significant changes on
the dynamics of the scalar field. Applying the numerical simulation, we calculate the present values
of the quantities as, $\Delta \phi (z=0) \simeq 0$, for all used values of $\beta$,
and $\Omega_V (z=0) \simeq$ 0.722, 0.734, and 0.758, respectively, for $\beta =$ 0.05, 0.1 and 0.2.\\

\begin{figure}
 	\includegraphics[width=8.5cm, height=3.5cm]{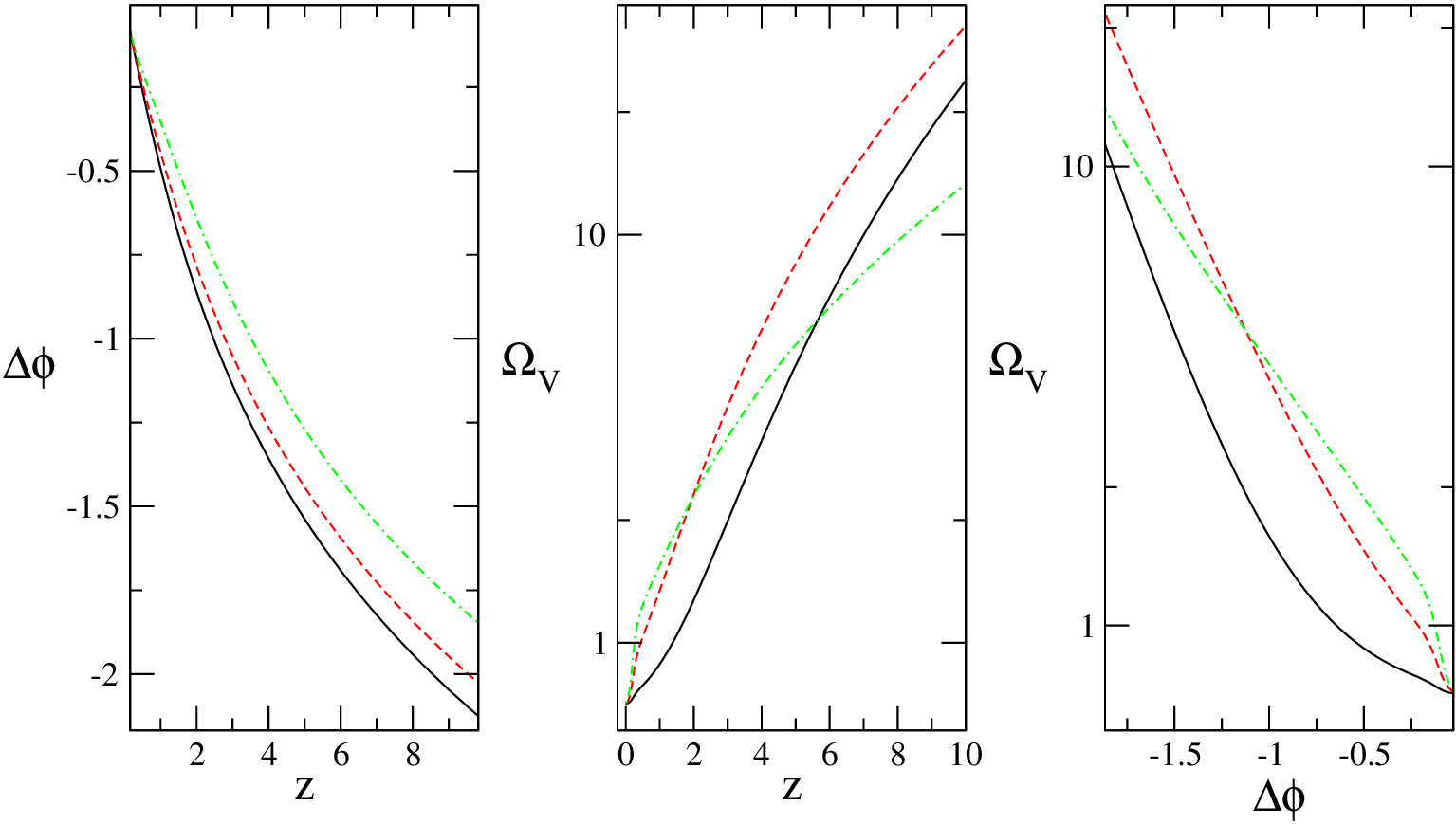}
     \caption{This displays an equivalent scalar field prescription for model III (see Eq. \ref{proposta3}).
Left panel: $\Delta \phi$ versus $z$. Middle panel: $\Omega_V$ versus $z$. Right panel: $\Omega_V$ versus $\Delta \phi$. Solid (black), dashed (red), and dot-dashed (green), stand for $\beta = 0.05, 0.1, 0.2$, respectively. We have taken $\Omega_{b0} = 0.05$, $\Omega_{dm0}=0.24$, for all the graphs.}
     \label{campo_model3}
\end{figure}

Figures \ref{campo_model2}, \ref{campo_model3}, respectively, show the corresponding scalar field descriptions
for the models II and III. Just like the dynamics associated with model I,
we note that, for $z \lesssim 2$, the particle production rate for model II, characterized by the parameter
$\beta$, has no significant influence on the variation of the field with respect to the redshift.
But, we notice a huge sensitivity in the evolution of the scalar potential at low redshift
(see the middle panels of the figures \ref{campo_model2}, \ref{campo_model3})
appearing due to $\beta$ at low redshifts. For model II,
we note, $\Delta \phi (z=0) \simeq 0$, irrespective of the value of $\beta$,
and $\Omega_V (z=0) \simeq$ 0.734, 0.829, and 0.948, respectively for $\beta =$ 0.01, 0.05 and 0.1.
For model III, the numerical results at low redshift for different values for the parameter
$\beta$ do not influence the dynamics of scalar field in a significant way. We note that,
$\Delta \phi (z=0) \simeq 0$ and $ \Omega_V (z=0) \simeq 0.71$ for all values of $\beta$ employed.

\section{Effects on the cosmic microwave background}\label{sec:cmb}

In this section, we explore the impact of a continuous matter creation process on the anisotropies of the
cosmic microwave background (CMB). Let us follow the treatment and the notation
as in \cite{cmb_1}, adopting a synchronous gauge. In this case, the line element of the linearly
perturbed FLRW metric can be written as

\begin{eqnarray}
ds^2 = -dt^2 + a^2(t)(\delta_{ij} + h_{ij})dx^idx^j,
\end{eqnarray}
where $h_{ij}$ represent the metric perturbation. Here, we will restrict ourselves to the scalar modes $h$ and $\eta$
of the metric perturbations, where $h$, $\eta$, are respectively the trace and traceless parts of $h_{ij}$, which in Fourier space are defined as \cite{cmb_1}

\begin{eqnarray}
h_{ij}(x,\tau) = \int d^3k \exp^{i\vec{k} \cdot \vec{x}} \Big[ \hat{k}_i \hat{k}_j h(\vec{k},\tau) + \nonumber \\ \Big(k_i k_j - \frac{1}{3} \delta_{ij}\Big) 6\eta(\vec{k},\tau) \Big], \, \,\,\,\,\, \vec{k} = k \hat{k}.
\end{eqnarray}
The conservation of energy-momentum is a consequence of the Einstein equations. Let
$w_c = p_c/\rho_{dm}$, describes the equation of state associated with matter creation.
Then the perturbed part of energy-momentum conservation equations

\begin{eqnarray}
T^{\mu \nu}_{;\mu} = \partial_{\mu}T^{\mu \nu} + \Gamma^{\nu}_{\alpha \beta} T^{\alpha \beta} + \Gamma^{\alpha}_{\alpha \beta} T^{\mu \beta} =0,
\end{eqnarray}
imply

\begin{eqnarray}
\label{delta_dm}
\dot{\delta}_{dm} = - (1+w_c) \Big(\theta_{dm} + \frac{\dot{h}}{2} \Big) - 3 H w_c \delta_{dm},
\end{eqnarray}
and

\begin{eqnarray}
\label{theta_dm}
\dot{\theta}_{dm} = - H(1-3w_c)\theta_{dm} - \left(\frac{\dot{w_c}}{1+w_c}\right) \theta_{dm},
\end{eqnarray}
where $\theta_{dm} = i k^i v^i_{dm}$, is the divergence of the peculiar velocity, and the equation of state $w_c$,
associated with the particle creation models are described in equations (\ref{w_c_1}), (\ref{w_c_2}), and (\ref{w_c_3}).
Note that, in absence of matter creation, i.e. when $w_c = 0$, we obtain the standard evolution
for the perturbations of dark matter. In the above equations, we have neglected the shear stress
of the dark matter which is always small because of its non-relativistic character.

\begin{figure}
 \includegraphics[width=6.9cm, height=5cm]{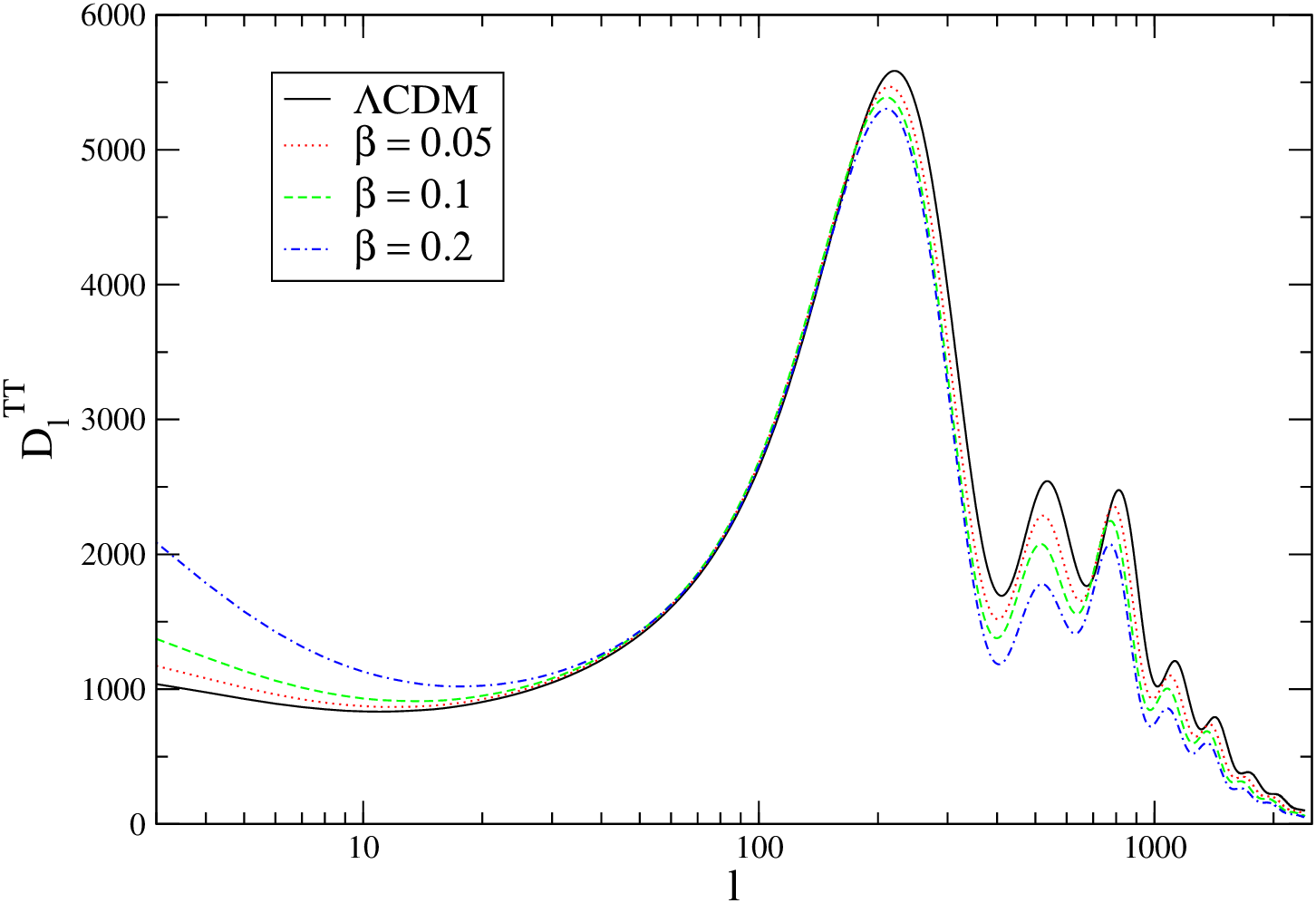}
     \caption{The figure shows the CMB TT power spectra represented by
     $D_l^{TT} = l(l+1)C_l / 2\pi$ $\mu K^2$ for model I (with three different values
     of $\beta$ labeled in the figure) and the flat $\Lambda$CDM.}
     \label{Cl_model1a}
\end{figure}

\begin{figure}
 	\includegraphics[width=6.9cm, height=5cm]{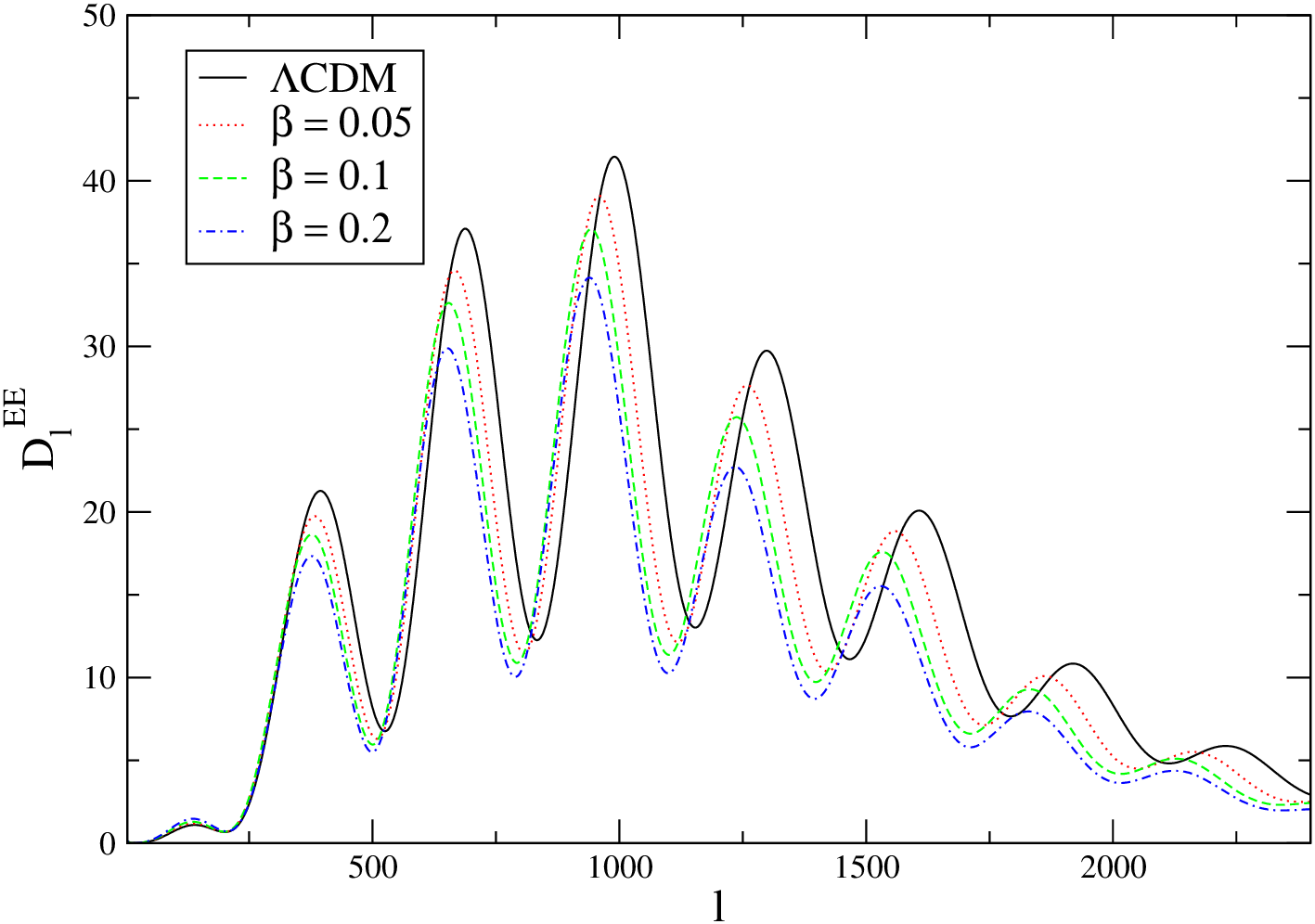}
     \caption{This figure shows the CMB EE power spectra represented by
     $D_l^{EE} = l(l+1)C_l / 2\pi$ $\mu K^2$ for model I (with three different values
     of $\beta$ labeled in the figure) and the flat $\Lambda$CDM.}\vskip 0.5 cm
     \label{Cl_model1b}
\end{figure}

\begin{figure}
 	\includegraphics[width=6.9cm, height=5cm]{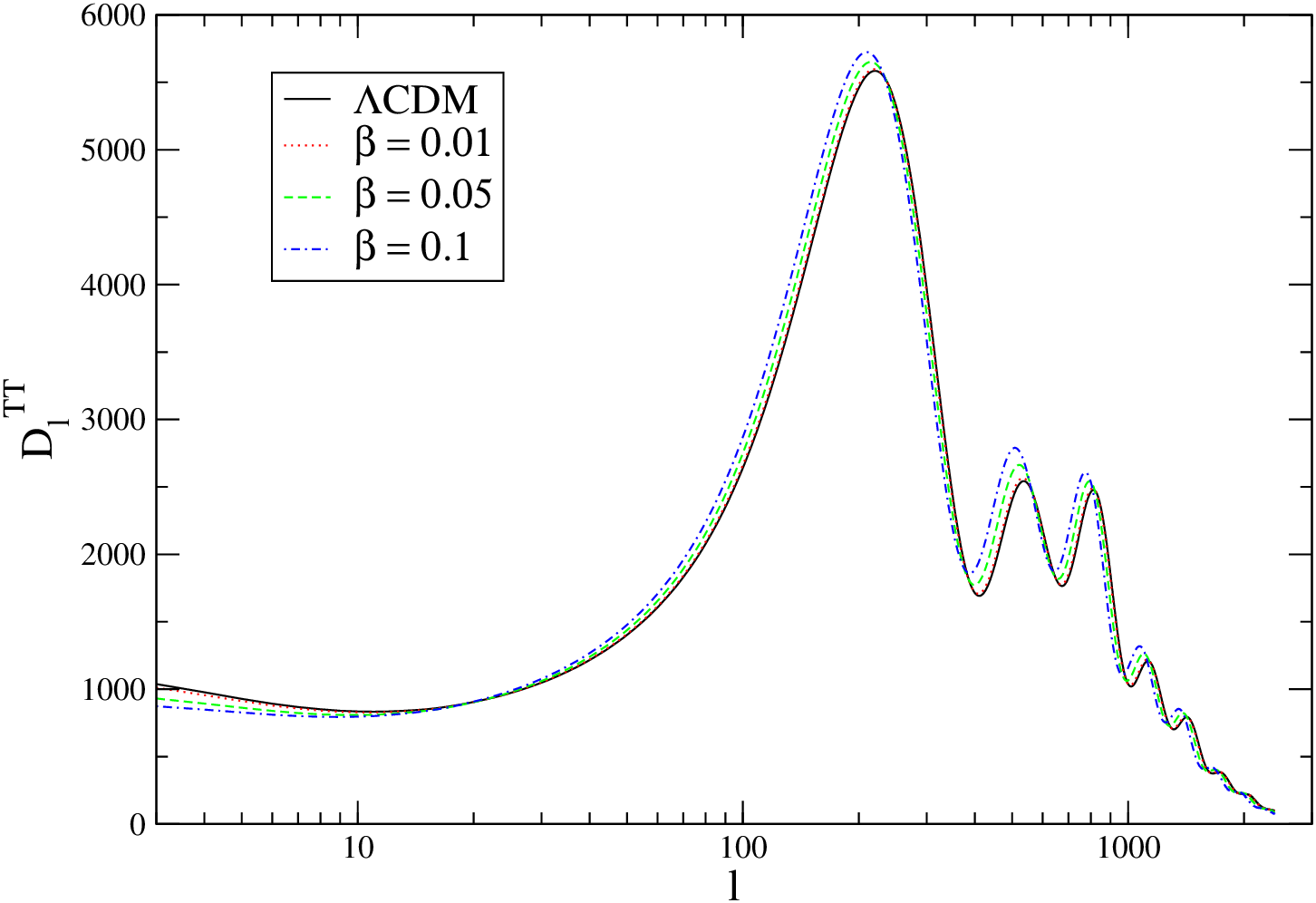}
 \caption{CMB TT power spectra represented by $D_l^{TT} = l(l+1)C_l / 2\pi$ $\mu K^2$ for model II (with three different values of $\beta$ labeled in the figure) and the flat $\Lambda$CDM.}
     \label{Cl_model2a}
\end{figure}

\begin{figure}
 	\includegraphics[width=6.9cm, height=5cm]{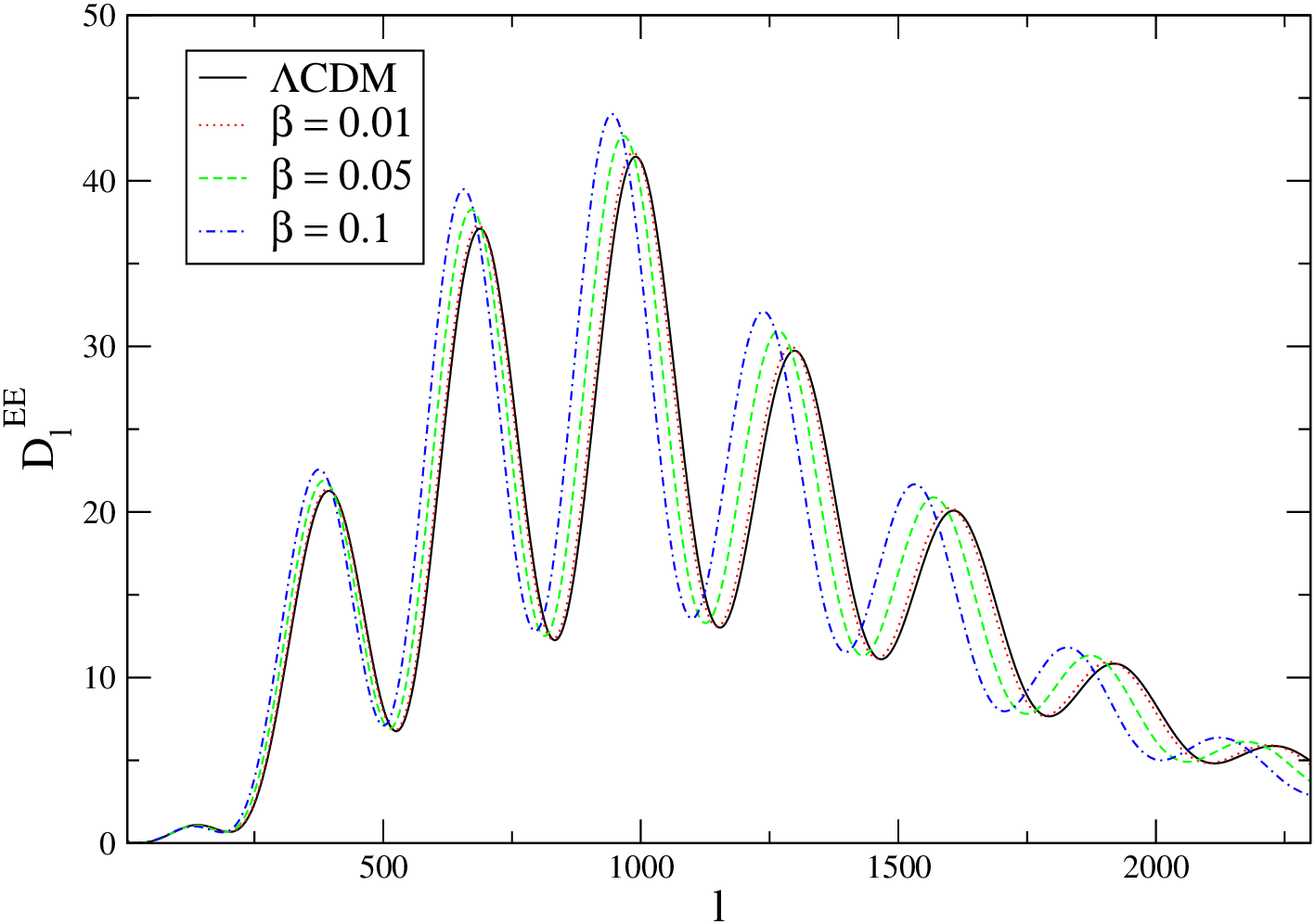}
     \caption{CMB EE power spectra represented by
     $D_l^{EE} = l(l+1)C_l / 2\pi$ $\mu K^2$ for model II (with three different values
     of $\beta$ labeled in the figure) and the flat $\Lambda$CDM.}\vskip 0.5 cm
     \label{Cl_model2b}
\end{figure}

\begin{figure}
 	\includegraphics[width=6.9cm, height=5cm]{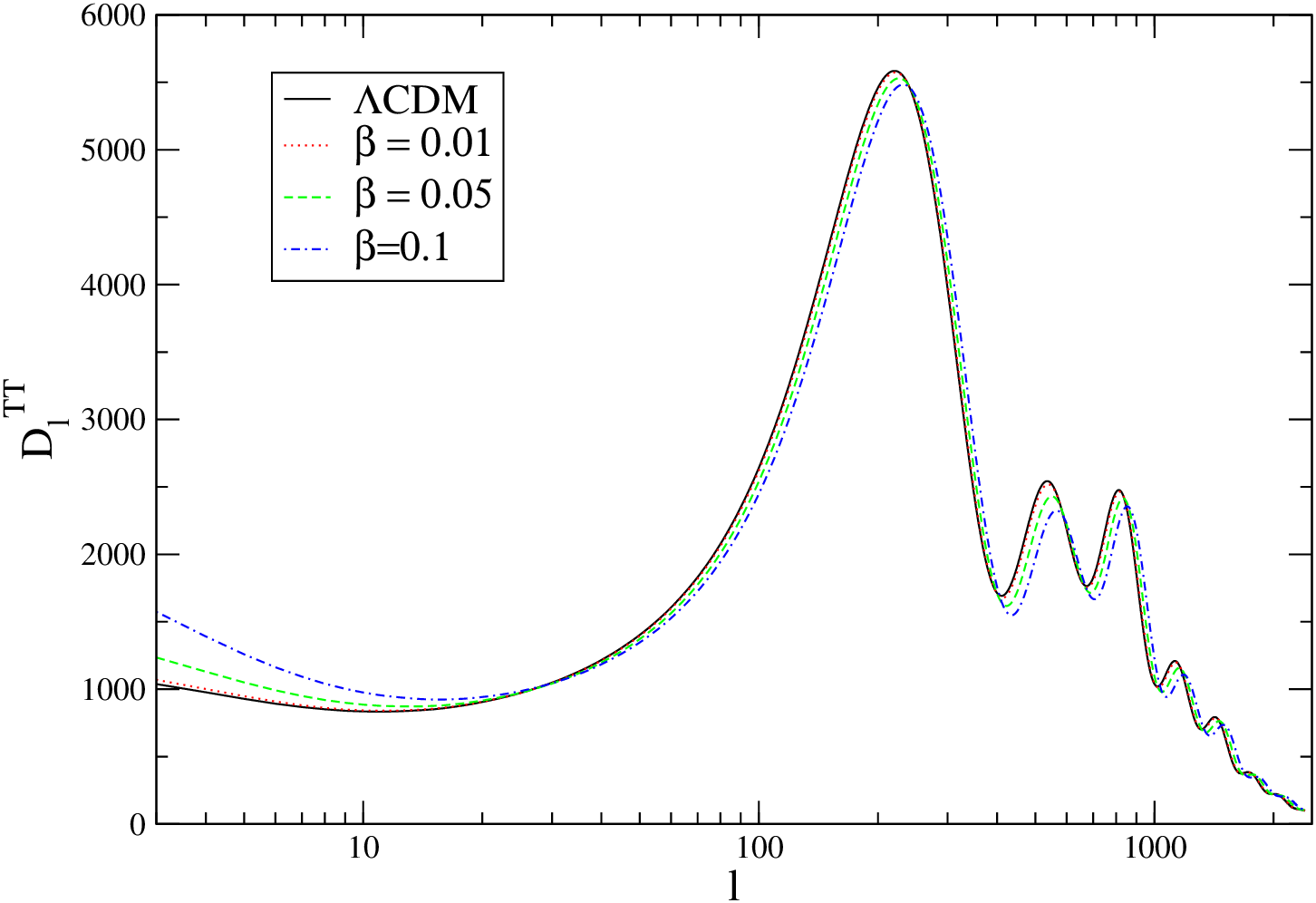}
\caption{CMB TT power spectra represented by $D_l^{TT} = l(l+1)C_l / 2\pi$ $\mu K^2$ for model III (with three different values of $\beta$ labeled in the figure) and the flat $\Lambda$CDM.}\vskip 0.5 cm
     \label{Cl_model3a}
\end{figure}

\begin{figure}
\includegraphics[width=6.9cm, height=5cm]{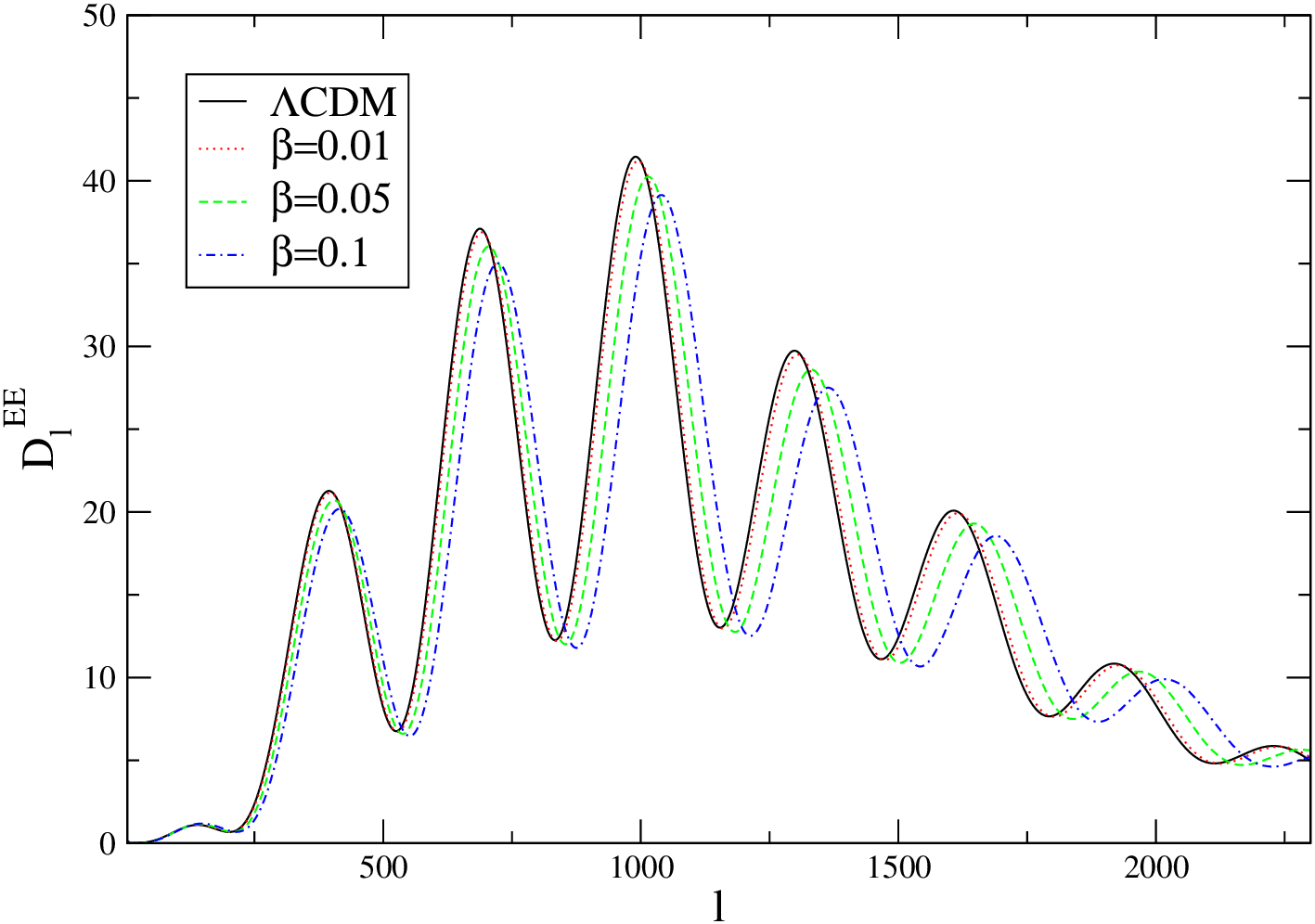}
\caption{CMB EE power spectra represented by $D_l^{TT} = l(l+1)C_l / 2\pi$ $\mu K^2$ for model III (with three different values of $\beta$ labeled in the figure) and the flat $\Lambda$CDM.}
     \label{Cl_model3b}
\end{figure}

Now, taking into account the  set of equations (\ref{delta_dm}), and (\ref{w_c_1}) --- (\ref{w_c_3}) that describe the evolution of the cold dark matter,
we modified CAMB \cite{camb} in order to see the effects of adiabatic particles production on CMB power spectra.

Figures \ref{Cl_model1a},  \ref{Cl_model1b}, respectively, show the CMB TT, CMB EE power spectra computed
for model I (with different values of $\beta$) and flat $\Lambda$CDM. 
Compared to the theoretical prediction of the flat $\Lambda$CDM model,
when the effects of particles production are introduced, we see a significant change in the amplitude and phase of the
acoustic peaks of the CMB, i.e. if $\beta$ increases, the changes increase.  In other words,
if the particle production increases, the changes in CMB power spectra increase.

Figures \ref{Cl_model2a}, \ref{Cl_model2b} respectively, show the CMB TT, CMB EE power spectra for
model II (with different values of $\beta$) and flat $\Lambda$CDM model.
Similarly, in figures \ref{Cl_model3a}, and \ref{Cl_model3b}, we respectively, plot the CMB TT and CMB EE power spectra for
model III (with different values of $\beta$) in compared to the flat $\Lambda$CDM model.
Note that, model II has similar behavior to the flat $\Lambda$CDM model in all angular scales on the CMB TT power spectrum, for all values of the parameter
$\beta$ compatible with the thermodynamic analysis. For $\beta \geq 0.05$, significant deviations on
CMB EE power spectrum have been found in compared to the theoretical prediction of the flat $\Lambda$CDM. 
Like model II, model III also shows a similar behavior to the $\Lambda$CDM model, but this applies only to small angular scales ($l \geq 30$) on
the CMB TT power spectrum. In case of CMB EE power spectrum, model III presents a significant deviation\footnote{We note that, in figures \ref{Cl_model2a}, \ref{Cl_model2b}, \ref{Cl_model3a}, \ref{Cl_model3b}, the power spectra calculated for models II and III with $\beta = 0.01$ overlapped with the power spectra for flat $\Lambda$CDM.} in compared to $\Lambda$CDM model for $\beta \geq 0.05$. 

Therefore, the deviation from the standard evolution law $(1+z)^3$ of dark
matter, due to the introduction of the effects of matter creation, affects the CMB temperature power spectra
at all angular scales, especially at the low $l$ via the integrated Sachs-Wolfe effect due to the evolution
of gravitational potential. Thus, from the results shown here, it is evident that the dynamic nature
associated with the particle production rate has significant influence on the evolution of perturbations.
Models with $w_c =$ const, i.e. model I, presents a higher sensitivity on the CMB power spectra in compared
to dynamic models for $w_c$, i.e. models II and III.

\section{Om diagnostic for a phantom behavior via cosmological creation of particles}
\label{sec:OM}

At present, we have a large number of cosmological models in order to explain the current accelerating universe. As the number of models is very large and increasing with time, so we should be careful about the models.
Specifically, we must know how much they are sound in compared to the latest observational data we have.
This includes a lot of time and very careful data analysis to be confirmed whether the cosmological model is long lasting or not. Thus, in this situation, a model independent test will be very much useful which could rule out some cosmological models from the literature. The one which brings itself into the dark energy discussions is the $Om$ diagnosis \cite{SSS2008}, which for the FLRW universe is a combination of the Hubble rate $H (z)$ and the redshift, where this Hubble rate is a model independent quantity of whose value can be obtained from the luminosity distance $D_L$ from the following equation \cite{Starobinsky1998, HT1999, NC1999, SRSS2000, CN2000}

\begin{eqnarray}
H (z) & = & \left[\frac{d}{dz}\left(\frac{D_L (z)}{1+z}\right)\right]^{-1}.
\end{eqnarray}

So, once $H (z)$ is determined from the luminosity distance, $Om$ can be constructed. However, it is worth noting that, if for any dark energy model, if $Om$ is found to be same at different redshifts, then the dark energy is the cosmological constant. So, $Om$ can be used to distinguish several dark energy models from $\Lambda$CDM.
Further, for any dark energy model with un-evolving equation of state, the nature of dark energy could be known, that means it is possible to tell whether the dark energy is of quintessence or phantom. In order to proceed further and present some more interesting qualities in this diagnosis, let us define $Om$ as \cite{SSS2008}

\begin{eqnarray}
Om (z)&=& \frac{h^2 (z)-1}{(1+z)^3-1},\label{om1}
\end{eqnarray}
where
\begin{eqnarray}
h (z)&=& \frac{H (z)}{H_0}~.\nonumber
\end{eqnarray}
Now, in this flat space-time, we consider a non-interacting two fluid cosmological model where one fluid is dark energy
with un-evolving equation of state $\omega_d$, and
the other fluid is pressureless matter. Hence, the expansion history $H (z)$ can be written as

\begin{eqnarray}\label{om1.1}
\left(\frac{H(z)}{H_0}\right)^2 &=& \Omega_{m0} (1+z)^{3}+ \Omega_{d0} (1+z)^{3(1+\omega_d)},
\end{eqnarray}
and consequently, the expression for $Om (z)$ becomes

\begin{eqnarray}
Om (z) &=& \Omega_{m0}+ \Omega_{d0} \left[\frac{(1+z)^{3 (1+\omega_d)}-1}{(1+z)^3-1}\right],\label{om1a}
\end{eqnarray}
which leads to following results: If we consider the flat $\Lambda$CDM model, we have $\omega_d= -1$, and
hence, Eq. (\ref{om1a}) reduces to $Om (z)= \Omega_{m0}= $ constant, throughout the evolution of the universe dominated by
the cosmological constant and the pressureless matter. In other words, for any two redshifts $z_1$, $z_2$,
$Om(z_1)- Om(z_2)= 0$, i.e. $Om (z)$ gives a null test for $\Lambda$CDM. Further,
if $Om (z)= \Omega_{m0}$ for any cosmological model, then from (\ref{om1a}) we have $\omega_d= -1$, i.e. DE $ = \Lambda$. Thus,
we see that, DE $= \Lambda$ $\Longleftrightarrow$  $Om (z)= \Omega_{m0}$.
On the other hand, if $\omega_d> -1$ (quintessence), from Eq. (\ref{om1a}), one has
$Om (z)> \Omega_{m0}$. Similarly, when $\omega_d< -1$ (phantom), $Om (z)< \Omega_{m0}$.
In connection with this discussion, we would like to introduce some
noteworthy points in $Om$ in compared to
the statefinders $\{r, s\}$ \cite{Sahni2003} (another model independent test for cosmological models) defined as

\begin{eqnarray}\label{statefinders}
r&=& \frac{1}{aH^3} \frac{d^3 a}{dt^3},~~~s= \frac{r-1}{3 \left(q-\frac{1}{2}\right)}.
\end{eqnarray}

For flat $\Lambda$CDM, we get $\{r, s\}= \{1, 0\}$, and henceforth, for any two
redshifts $z_1 \neq z_2$, $r(z_1)- r(z_2)= 0$, leading to another null test for $\Lambda$CDM
as given by $Om (z)$ too. But, unlike
$r= \dddot{a}/aH^2$, which contains the third order derivative with respect to the cosmic
time, $Om$ depends only on the
expansion history ($H (z) $) of our universe, a first order derivative of the scale factor.
On the other hand, it can be
shown that, $Om$ at two different redshifts could differentiate between different
dark energy models without any need of $H_0$ or the matter density $\Omega_{m0}$. In order to
show our claim to be true, we consider the following expression

\begin{eqnarray}\label{om-new}
Om (z_1, z_2)= Om (z_1)- Om (z_2) \nonumber\\=  (1-\Omega_{m0}) \left[\frac{(1+z_1)^{3 (1+\omega_d)}-1}{(1+z_1)^3-1}-\frac{(1+z_2)^{3 (1+\omega_d)}-1}{(1+z_2)^3-1}\right]
\end{eqnarray}
which for the cosmological constant establishes the relation, $Om (z_1, z_2)= 0 \Longleftrightarrow$ DE $= \Lambda$;
$Om(z_1,z_2)> 0$ for quintessence; and $Om (z_1, z_2)< 0$ for the phantom fluid ($z_1< z_2$). So, to determine
the nature of dark energy, we do not need the present matter density as well as $H_0$.
Hence, it is more easy to reconstruct $Om$ from observations than the statefinders,
as well as, its next extension to higher order terms, known as cosmography \cite{Visser04, Visser05}.
Now, corresponding to our three phenomenological particle creation models,
we get three different expressions for $Om (z)$ as

\begin{eqnarray}
Om (z)&=& \frac{\Omega_{b0} (1+z)^3+ \Omega_{dm0} (1+z)^{3 (1-\beta)}+ \Omega_{\Lambda0}-1}{(1+z)^3-1},
\end{eqnarray}
for model I, and the expression

\begin{eqnarray}
Om (z)= \frac{1}{(1+z)^3-1} \Bigl[\Omega_{b0} (1+z)^3 \nonumber\\+  \Omega_{dm0} (1+z)^3 \exp\left(-3 \beta \int_0^z  \frac{\psi (\tilde{z})}{(1+\tilde{z})}d\tilde{z}\right)+ \Omega_{\Lambda0}-1\Bigr],
\end{eqnarray}
stands both for models II and III, where $\psi (z)$ for models II, III are given in Eqns. (\ref{gaII}), (\ref{gaIII}), respectively. Note that, if $\beta \longrightarrow 0$, model I can not be distinguished from $\Lambda$CDM, and if $\beta$ increases, model I starts deviating from the $\Lambda$CDM. That means, if matter creation increases, model I shows a significant deviation from $\Lambda$CDM as observed in CMB power spectra for increasing $\beta$. On the other hand, for models II, III, we find that, $0 \leq \exp\left(-3 \beta \int_0^z  \frac{\psi (\tilde{z})}{(1+\tilde{z})}d\tilde{z}\right) (= f (z)) \leq 1$, for all $z$, irrespective of $\beta$, and this quantity tends to $1$ as $z \longrightarrow 0$. So, at late-time, models II, III almost coincide with $\Lambda$CDM. Further, we notice that, if $\beta$ increases, $f (z)$ decreases, and on the other hand, if $\beta$ decreases, $f (z)$ increases, but in all cases, this quantity is always bounded in [0, 1]; furthermore, for $\beta= 0$, it is readily seen that the models II, III coincide with $\Lambda$CDM.
\begin{figure}
 	\includegraphics[width=6.9cm, height=5cm]{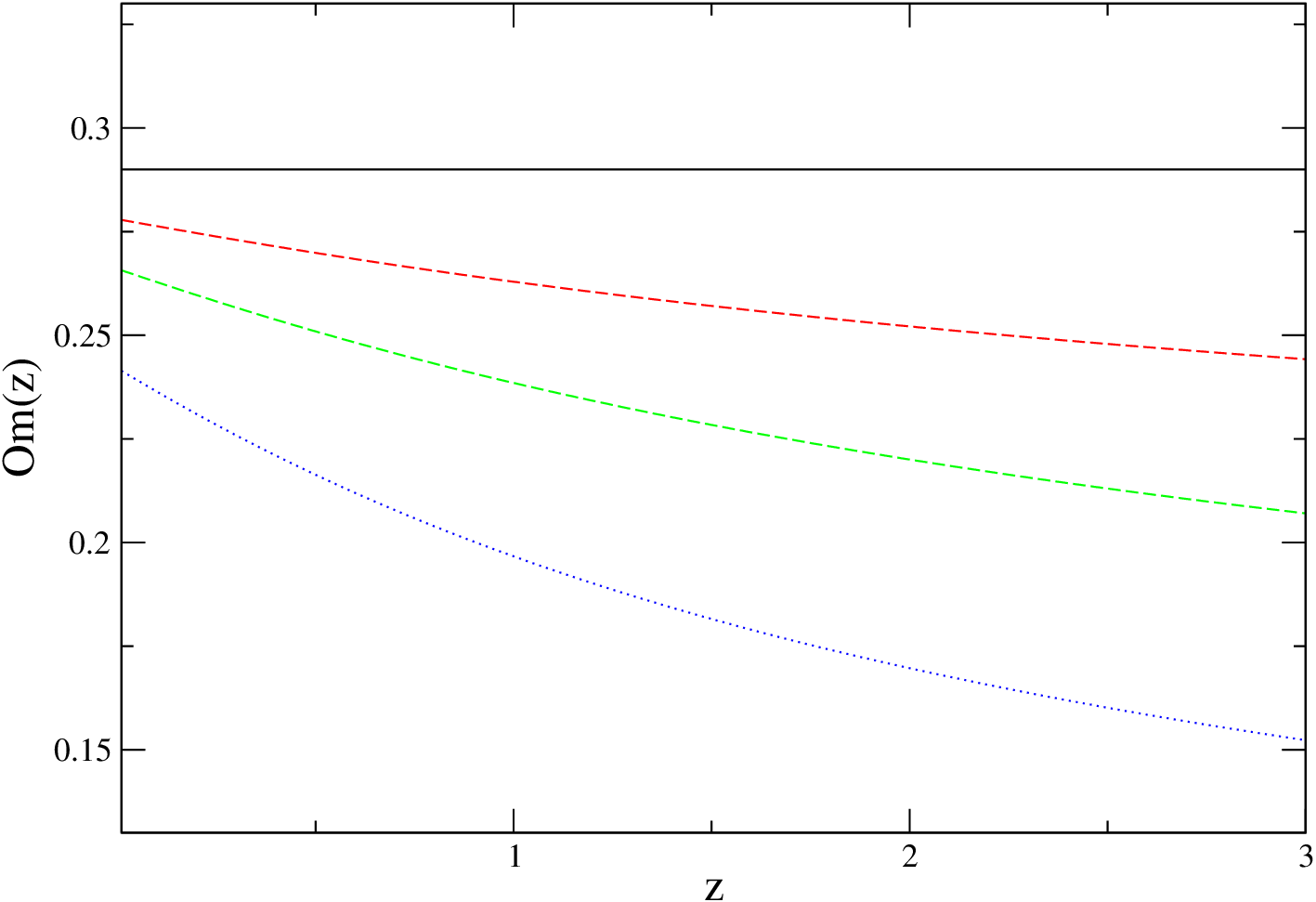}
     \caption{The figure shows the evolution of $Om(z)$ for the model I (see Eq. \ref{proposta1}) in compared to the flat $\Lambda$CDM, where the solid (black), dashed (red), dot-dashed (green), and dot (blue) lines respectively represent the flat $\Lambda$CDM, and the model I with $\beta= 0.05$, $\beta= 0.1$, $\beta= 0.2$. While drawing the graphs, we have taken $\Omega_{b0} = 0.05$, $\Omega_{dm0}=0.24$.}\vskip 0.5cm
     \label{om_model1}
\end{figure}

\vskip 0.9cm

\begin{figure}
 	\includegraphics[width=6.9cm, height=5cm]{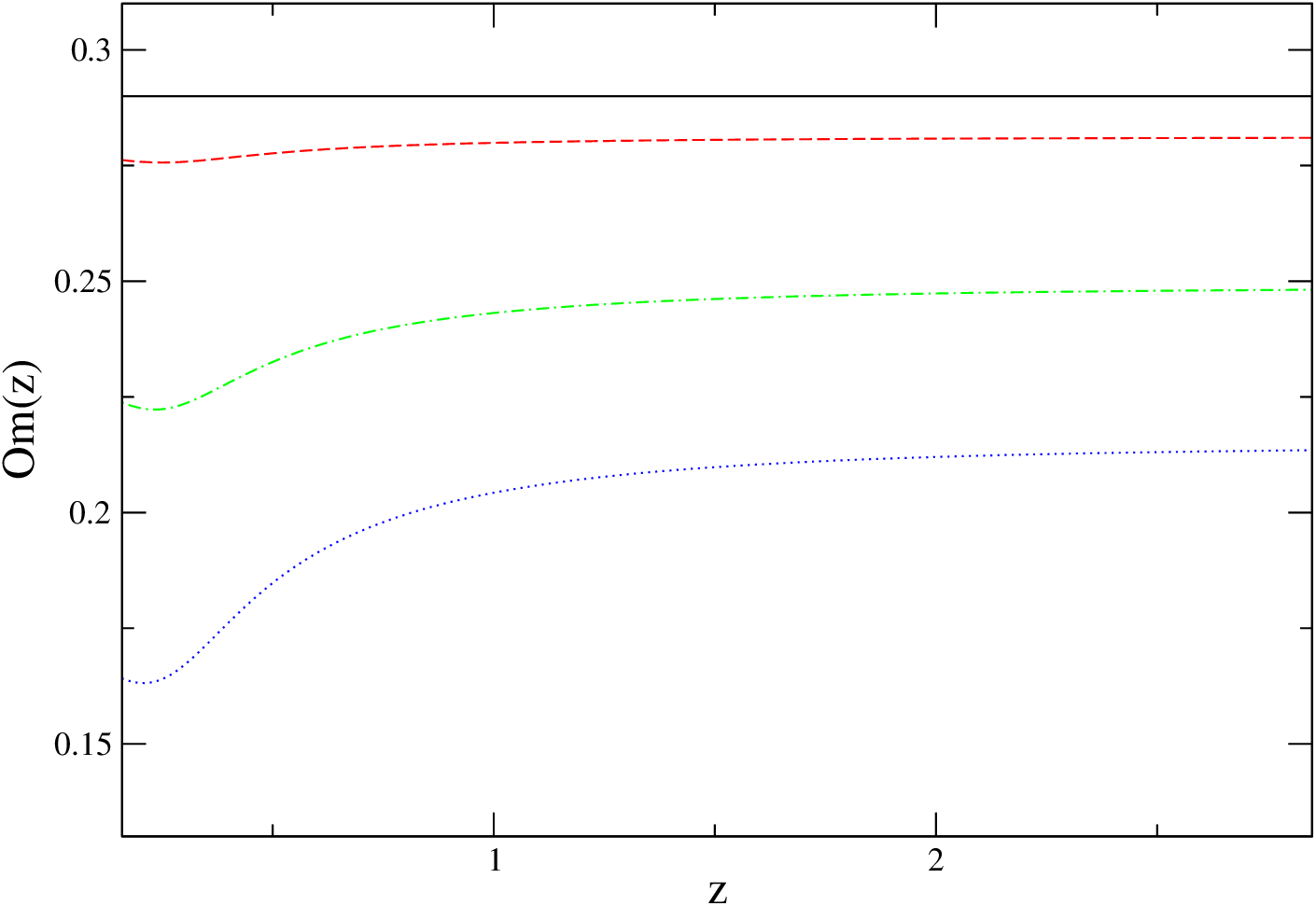}
     \caption{The figure depicts the evolution of $Om$ for model II (see Eq. \ref{proposta2}) and flat $\Lambda$CDM, where the solid (black), dashed (red), dot-dashed (green), and dot (blue) lines respectively represent the flat $\Lambda$CDM, and the model II with $\beta= 0.05$, $\beta= 0.1$, $\beta= 0.2$. Here, we have taken $\Omega_{b0} = 0.05$, $\Omega_{dm0}=0.24$.}
     \label{om_model2}
\end{figure}

\begin{figure}
 	\includegraphics[width=6.9cm, height=5cm]{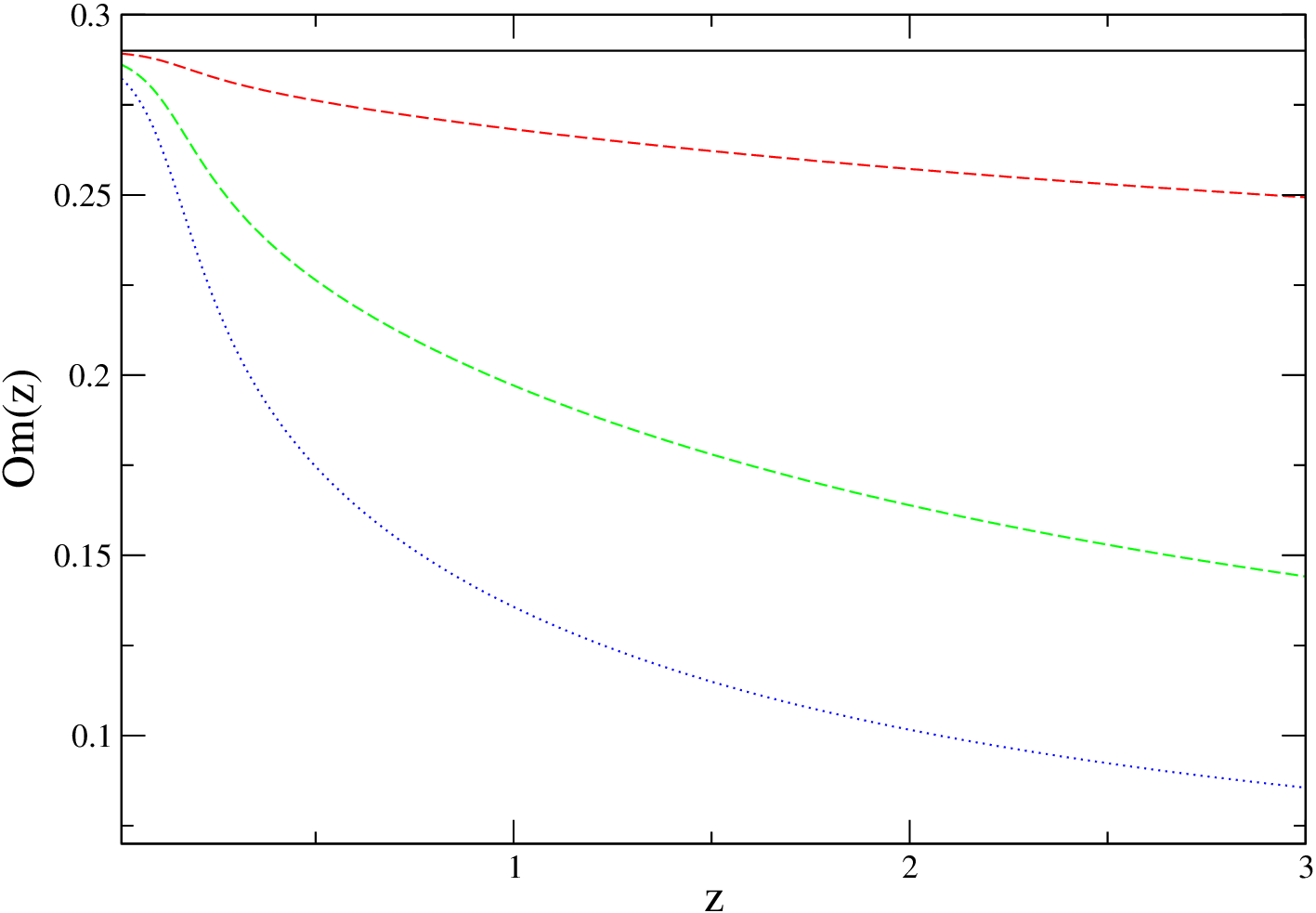}
     \caption{Similar to figures \ref{om_model1}, \ref{om_model2}, this shows the evolution of $Om$ for model III (see Eq. \ref{proposta3}) and flat $\Lambda$CDM, where the solid (black), dashed (red), dot-dashed (green), and dot (blue) lines respectively represent the flat $\Lambda$CDM, and the model III with $\beta= 0.05$, $\beta= 0.1$, $\beta= 0.2$. Here, we have taken $\Omega_{b0} = 0.05$, $\Omega_{dm0}=0.24$.}
     \label{om_model3}
\end{figure}

In figures \ref{om_model1}, \ref{om_model2}, and \ref{om_model3}, we show
the evolution of the $Om$ functions corresponding to models I, II, III, respectively.
Note that, all the models pass the $Om$ test confirming their phantom behavior, i.e. throughout the evolution of the universe,
the value of $Om$ corresponding
to each model is always lower than the value of $Om$ for $\Lambda$CDM. 
Therefore, the evolution of $Om$  presents a significant variation that depends on the particle production rates.

Figure \ref{om_model1} shows $Om$ diagnosis for model I, where note that,
when $z$ decreases, $Om$ increases, but always lower than the $Om$ value taken
by $\Lambda$CDM. The point which should be noted is that, if $\beta$ increases
(in other words, if matter creation increases), $Om$ decreases.
That means $Om$ is inversely related with the matter creation for the model I.

Figure \ref{om_model2} shows the variation of $Om$ for model II. Here, we notice a similar behavior to model I in low redshifts (see figure \ref{om_model1}). In this case, note that for $z> 1$, $Om$ is almost constant with the evolution of the universe. Figure \ref{om_model3} shows the $Om$ diagnostic applied to the model III. In this scenario some properties which should be discussed. It is again true that, model III is inversely related with the $Om$ function. The most important point we notice in this case is that, as $z\longrightarrow 0$, $Om$ rapidly increases, and approaches towards the $\Lambda$CDM model. Model III with $\beta= 0.05$ almost coincides with $\Lambda$CDM, although for $\beta= 0.1$, $\beta= 0.2$, the difference is very slight with $\Lambda$CDM. Thus, at low redshift, with small creation rate, model III almost coincides with $\Lambda$CDM.

\section{Thermodynamic analysis}
\label{sec:thermodynamics}

\noindent For a cosmological model to be worthy of further
investigations, it is not only necessary for the model to be fitted with the current astronomical data,
but also, the model should be consistent with the thermodynamics of the universe.
The latter asserts that, the entropy of an isolated
macroscopic system can never diminish, and it has to be
concave at least during the last stage of approaching
thermodynamic equilibrium \cite{callen}.
\\


\noindent In the context of an ever expanding FLRW universe, the
full version of the thermodynamic law implies that the entropy of
the apparent horizon plus that of matter and fields enclosed by it
must fulfill $\, {\rm d}S/{\rm d}a \geq 0 \,$, for any scale factor,
the so-called generalized second law of thermodynamics (GSLT), and ${\rm
d}^{2}S/{\rm d}a^2 \leq 0 \,$, as $\, a \rightarrow \infty$
\cite{grg-nd1}.
\\

\noindent As demonstrated by Bak and Rey \cite{bak-rey}, cosmic
apparent horizons are  endowed with an entropy proportional to the
horizon area, much in the same way as black holes have an entropy
that varies as the area of its event horizon. Unlike cosmic event
horizons, the apparent horizon\footnote{In connection with the apparent horizon, we mention that as surface gravity and temperature are analogous \cite{JP, TP1, TP2}, and since apparent horizon does not include any null surface hence no surface gravity, so temperature does not make any sense with the apparent horizon.} always exists and coincides with
the former in case of a final de Sitter expansion.
\\

\noindent Thus, in accordance with the GSLT, the total entropy $S$
is contributed by two terms: the entropy of the apparent horizon,
$S_h=k_B \, {\cal A}/(4 \ell^{2}_{pl}), $ with ${\cal A} = 4 \pi
r^{2}_{h}$ being the area of the horizon, and the entropy of the
fluids and fields enclosed by it (pressureless matter in the case
at hand). Here, $r_{h}=(\sqrt{H^2+k/a^2})^{-1}$ denotes the radius
of the apparent horizon. In our case, for a spatially-flat universe
($k= 0$), the latter coincides with the Hubble horizon radius,
$H^{-1}$. Thus, GSLT asserts that ${\rm d}S/{\rm d}a = {\rm
d}(S_{h} + S_{m})/{\rm d}a \geq 0$.
\\  \

\noindent Recalling the above expression for the horizon entropy,
one has $\, {\rm d}S_{h}/{\rm d}a= -2 \pi k_B ({\rm d}H/{\rm
d}a)/(\ell^{2}_{pl} H^3) $. Now, for the model I, we have
\begin{eqnarray}
\label{Derivada_Hz_1}
 \frac{{\rm d}H}{{\rm d}a}= - \frac{3}{a}\frac{H^2_0}{2H}
 \left\{ (1 - \beta)  \Omega_{dm0} a^{-3(1 - \beta)}
 + \Omega_{b0} a^{-3} \right\},
\end{eqnarray}
and for the models II and III, we have
\begin{eqnarray}
\label{Derivada_Hz_2}
 \frac{{\rm d}H}{{\rm d}a}= - \frac{3}{a}\frac{H^{2}_{0}}{2H} \Bigl\{ [1 - \beta g(a)] \, \Omega_{dm0} a^{-3}
 \exp \left(3\beta \int_{1}^{a} {\frac{g(\tilde{a})}{\tilde{a}} {\rm d}\tilde{a}}\right)
 \, \nonumber\\+ \, \Omega_{b0} \, a^{-3}\Bigr\}
\end{eqnarray}
where $g(a)$ can be found in equations (\ref{gaIIa}) and (\ref{gaIIIa}).

We notice that, for model I, $dS_h/da > 0$, if $\beta< 1$. On the other hand, for models II and III, ${\rm d}S_h/{\rm d}a$ will be
non-negative, provided that, $\, g(a) \beta \leq 1 $, and $\beta \leq 1$.\\

\noindent Now, for the variation of the entropy of dust matter, it
suffices to realize that every single particle contributes to the
entropy inside the horizon by a constant bit, say $k_B$. Then
$S_m =k_B 4\pi n r^{3}/3$, where the number density of dust
particles, $n = n_{b} + n_{dm}$, obeys the conservation equation
(\ref{Eq:nbalance}).
\\  \

\noindent Thus,
\begin{eqnarray}
\label{variacion_entropia1} \frac{{\rm d}S_m}{{\rm d}a} =
  \frac{4 \pi k_B}{H^3}   \left[ -\frac{3}{H} \frac{{\rm d}H}{{\rm d}a} + \frac{{\rm d}n}{{\rm d}a} \right].
\end{eqnarray}
In  case of model I, this quantity is guaranteed to be
positive-semidefinite, if $\, 0 < \beta < 1 \,$, and the present
number of dark matter particles exceeds that of baryons, which is a very
reasonable assumption.
\\  \

\noindent For models II and III, one has
\begin{eqnarray}
\label{variacion_entropia2} \frac{{\rm d}n}{{\rm d}a}=-3 n_{b0} \,
a^{-4} \Bigl\{1 \nonumber\\+ \frac{n_{dm0}}{n_{b0}} \exp \left( 3\beta
\int_{1}^{a} \frac{g(\tilde{a})}{\tilde{a}}\, {\rm d}\tilde{a}
\right) \left[\beta g(a) - 1 \right] \Bigr\},
\end{eqnarray}
whence ${\rm d}S_{m}/{\rm d}a$ is assured to be
positive-semidefinite if the above assumption is met, and $\beta
g(a) < 1$.
\\

\noindent We next consider the sign of ${\rm d}^2 S_{h}{\rm d}a^2
+ {\rm d}^2 S_{m}/{\rm d}a^{2}$, as $a \rightarrow
\infty$. From ${\rm d}S_h/{\rm d} a= -\frac{2 \pi
k_B}{{\ell^{2}}_p} ({\rm d} H/{\rm d} a)/H^3 $, it follows,
\begin{eqnarray}
  \frac{{\rm d}^{2} S_{h}}{{\rm d} a^2}= - \frac{2 k_B \pi}{\ell^{2}_{pl}H^4}
  \left[ H\frac{{\rm d}^{2} H}{{\rm d}a^{2}} -3 \left(\frac{{\rm d}H}{{\rm d}a} \right)^{2}\right].
\end{eqnarray}
\\

\noindent Since as noted above,  $g(a)/a \rightarrow 0$, when $a
\rightarrow \infty$, one follows $\Omega_{dm}(a \rightarrow
\infty) = 0$, and $H(a \rightarrow \infty)=H_0
\sqrt{\Omega_{\Lambda}}$ (see Eqs. (\ref{Eq:Hz})
and (\ref{g_a})). Equation (\ref{Derivada_Hz_2}) implies
that,  if $\beta g(a) < 1$, then ${\rm d} H(a \rightarrow
\infty)/{\rm d} a$ $\rightarrow 0$, so it follows that $\, H ({\rm
d}^{2} H/{\rm d} a^2) -3({\rm d} H/{\rm d}a)^2 \rightarrow 0 \,$,
in that limit. The same procedure can be applied for model I.
\\

\noindent To discern the sign of ${\rm d}^2 S_{m}/{\rm d}a^2$, it
suffices to recall that, ${\rm d}H(a \rightarrow \infty)/{\rm d}a
\rightarrow 0$; then ${\rm d}S_{m}(a \rightarrow \infty)/{\rm
d}a=0$, and on the other hand, ${\rm d}S_{m}(a < \infty
)/{\rm d}a < 0$. Combining the last two expressions, we find that ${\rm
d}S_{m}/{\rm d}a$ tends to zero from below, i.e. ${\rm
d}^{2} S_{m}(a \rightarrow \infty)/{\rm d}a^{2} < 0$. Note that,
the conditions for models II and III, in agreement with the thermodynamics,
imply that $\Gamma/3H = \beta g(a) \leq 1$. For model I, the condition
is just $\beta \leq 1$. Altogether, when $a \rightarrow \infty$
one has ${\rm d}^{2} S_{h}/{\rm d}a^{2} + {\rm d}^{2} S_{m}/{\rm
d}a^{2} \leq 0$, as expected. In other words, for the three
phenomenological models considered in this work, our universe
behaves as an ordinary macroscopic system \cite{grg-nd2}; i.e. it
eventually tends to thermodynamic equilibrium. Thus, the same
condition that guarantees the fulfillment of the GSLT ensures that
the total entropy is concave (${\rm d}^{2}S/{\rm d}a^{2} <0$), at
late times.\\

\noindent Altogether, the observational constraints on the
three particle creation models introduced in \cite{NP2015},
are consistent with the requirement that the universe attains a thermodynamic equilibrium
(i.e. the state of maximum entropy represented by the final de Sitter expansion) in the long run.

\section{Concluding remarks}
\label{sec:conclusions}

\noindent Cosmological models driven by the gravitational adiabatic particle production have
been intensively investigated as a viable alternative to the $\Lambda$CDM cosmology \cite{ccdm_model}.
Recently, Nunes and Pav\'{o}n \cite{NP2015} explored the possibility that the EoS
determined by recent observations \cite{Planck2015, Planck2013, Rest, Xia, Cheng, Shafer} is in reality an effective EoS that results from adding the negative EoS, $w_c$ (associated to the particle production pressure from the gravitational field acting on the vacuum) to the EoS of the vacuum itself, $w_{\Lambda}$.
In the present paper, we investigate the cosmological consequences of this  scenario. We have established a canonical scalar field description for the parametric models, and analyzed the dynamic behavior of the variation of the scalar field, $\Delta \phi$, and the potential $V (\phi)$, with the evolution of our universe governed by the matter creation process. Further, we have evaluated the  matter creation effects on the CMB power spectra, and compared the results with the theoretical prediction for the $\Lambda$CDM model. We found that the matter creation phenomena can significantly interfere with the amplitude of perturbations in CMB level, with strong indication from the dynamic nature of the EoS associated with the particles production rate (see Eqns. (\ref{w_c_1}) - (\ref{w_c_3})\,).\newline

On the other hand, we have introduced a model independent test $Om (z)$ which distinguishes several cosmological models from $\Lambda$CDM by providing a null
test for the dark energy to be the cosmological constant $\Lambda$, for which $Om (z)= \Omega_{m0}$. The test is also used to differ several cosmological models by indicating their nature (phantom or quintessence). We found that, the function $Om (z)$ for the three models (see figures \ref{om_model1}, \ref{om_model2}, \ref{om_model3}) confirms their phantom behavior as constrained by recent observational data (see last paragraph of section \ref{sec:models}). We notice that, the parameter $\beta$ plays an important role to deviate all models from $\Lambda$CDM, which is very clear from their figures.
Furthermore, it is worthy to mention that, model I exhibits similar behavior to both $Om$ and CMB power spectra, that means if $\beta$ increases, the deviation from $\Lambda$CDM becomes dominant both in CMB power spectra and by $Om$ diagnosis. In case of models II, III, we see that if $\beta$ increases, the deviation from $\Lambda$CDM becomes clearer both in CMB TT power spectra and by $Om$ functions. But, CMB EE power spectra do not present any significant variation for different $\beta$.\newline

Further, we examined GSLT for the three models, and found that, the observational constraints obtained for the three parametric models proposed in \cite{NP2015} are in agreement with the requirement that the universe attains a thermodynamic equilibrium in its long run. We have described the behavior of the models in a compact way in table \ref{table-comparison}.\newline

\begin{table*}
\caption{The table shows a comparative study of all three particle creation models taken into account their response to CMB power spectra, $Om$ diagnosis, and with thermodynamics. Here, we mean by SD $=$ Significant deviation, and NSD $=$ No significant deviation, in compared to flat $\Lambda$CDM model. }
\begin{tabular}{c c c c c c}
\hline
\hline
Models &$\beta$ & CMB TT& CMB EE& Prediction by $Om$ & GSLT \\
\hline
\hline
Model I& If increases &SD& SD& Always Phantom with SD& Holds, provided $\Gamma/3H \leq 1$\\
Model II& If increases &NSD& SD& Always phantom with SD & Holds, provided $\Gamma/3H \leq 1$\\
Model III& If increases &NSD& SD& Always phantom with SD for $z\gtrsim 0.5$& Holds, provided $\Gamma/3H \leq 1$\\
\hline
\hline
\end{tabular}
\label{table-comparison}
\end{table*}

Finally, the parametric models for particle productions different from the present models are also worth exploring.
But, the most important point which we would like to note that,
the cosmology of particle creation awaits until the nature of the created dark matter particles is discovered.

\section*{Acknowledgments}

R.C.N. acknowledges the financial support from CAPES Scholarship Box 13222/13-9.
SP acknowledges the Department of Atomic Energy (DAE), Govt. of India, for a Post-Doctoral grant (File No. 2/40(60)/2015/R\&D-II/15420).
SP also acknowledges the Department of Mathematics, Jadavpur University, where the work was initiated and some of the parts were carried out.
The authors are very grateful to Diego Pav\'on for his useful comments.
The authors are also very grateful to the anonymous referee for many illuminating comments which improved
the manuscript considerably. Lastly, we must acknowledge the assistant editor for pointing out some setting problems
in the figures.


\end{document}